\documentclass[aps,prl,twocolumn,showpacs,groupaddress,floatfix]{revtex4}
\usepackage{subfigure}
\usepackage{graphicx}
\usepackage{dcolumn}
\usepackage{epsfig}
\usepackage{amssymb}
\usepackage{amsmath}
\usepackage{rotating}
\usepackage{color}

\begin{document}


\title{Giant amplification of interfacially driven transport by hydrodynamic slip: diffusio-osmosis and
beyond}

\author{Armand Ajdari$^{1,2,3}$}
\author{Lyd\'eric Bocquet$^{4}$}
 \affiliation{$^{1}$Physico-Chimie Th\'eorique, UMR 7083 CNRS-ESPCI, 10 rue
Vauquelin, 75231 Paris, France\\
$^{2}$Department of Mathematics, MIT, Cambridge MA 02139, U.S.A.\\
$^{3}$DEAS, Harvard University, 29 Oxford Street, Cambridge MA 02138, U.S.A.\\
$^{4}$LPMCN, Universit\'e Lyon 1, UMR 5586 CNRS, 69622 Villeurbanne, France}
\date{submitted to Phys. Rev. Lett., January 23, 2006-- Resubmitted version of  april 21.}

\begin{abstract}
We demonstrate that "moderate" departures from the no-slip hydrodynamic boundary condition  (hydrodynamic slip lengths in the nanometer range) 
can result in a very large enhancement - up to two orders of magnitude- of most interfacially driven transport phenomena.
We study analytically and numerically the case of neutral solute diffusio-osmosis in a slab geometry to account for non-trivial couplings between interfacial structure and hydrodynamic slip.    
Possible outcomes are fast transport of particles in externally applied or self-generated gradient, and flow enhancement in nano- or micro-fluidic geometries. 
\end{abstract}
\maketitle

{\em Introduction -}
The advent of "microfluidics" and "nanofluidics" 
has motivated
the current great interest in understanding, modeling and generating
motion of liquids in artificial or natural networks of ever more tiny channels or pores \cite{Squ05}. 
Because of the huge increase in hydrodynamic resistance that comes with downsizing,
two avenues for moving efficiently fluid at such scales have been revisited.
Both rely on phenomena originating at the solid/liquid interface to take advantage of the increase of surface to volume ratio. 

The first one is
the {\em generation} of flow within the interfacial structure by application of a macroscopic gradient.  Electro-osmosis, i.e. flow-generation by an electric field, is the best known example which is commonly used in microfluidics\cite{Sto04}. But other surface-driven phenomena fall in the same category, such as diffusio-osmosis and thermo-osmosis where gradients of solute concentration and of temperature are used
to induce solvent flow \cite{Hun,And89}. Their phenomenology is usually
best described by an "effective slip" velocity, which quantifies the motion of the fluid with respect to the solid due to shearing forces in the usually thin interfacial layer \cite{And89}.

The second is the amplification of pressure-driven flow by surfaces such that the fluid hydrodynamically "slips" on the solid, as usually quantified by the so-called slip length $b$ \cite{ref} (the distance {\em within} the solid at which the flow profile extrapolates to zero). Recent efforts in this domain 
have concluded that with a clean "solvophobic" surface chemistry one can reach slip lengths up to a few ten nanometers \cite{Cot05}, but not much more unless topographic structures are specifically engineered \cite{Roth}.

In this paper, generalizing a point recently made for electro-osmosis \cite{Chu,Sto04,Jol04}, we argue that these two strategies can actually  be synergetically combined,
yielding strongly enhanced interfacial driven flows on "solvophobic" surfaces.
More quantitatively, we argue that an actual "hydrodynamic slip" increases
the "effective slip velocity",  which controls all manifestations of the interfacially driven-phenomena, 
by a factor $(1+b/L)$, where $b$ is the hydrodynamic slip length and $L$ a measure of the interfacial thickness. This ratio can be {\it of order ten to hundreds} in realistic situations, so that the enhancement described here can be very large. This synergy may lead to more efficient transduction of
electrical, chemical or thermal energy into mechanical work in micro-devices.

Beyond the nano/microfluidic interest in moving fluids in tiny solid structures, our considerations also apply to the reciprocal interfacially driven motion of solid particles in
solution. We thus predict enhancement of electrophoresis, diffusiophoresis, and thermophoresis (induced respectively by gradients in electric potential, concentration of solutes
and gradients of temperature) when solvophobic particles are dispersed in solution.
Our analysis may also be of relevance to the "swimming" of artificial or natural organisms
by self-generation of such gradients \cite{And89,Lam,Pax,Gol05}. 

To exhibit the physics at work, we first focus on diffusio-osmosis  
with a single neutral solute species, in the simplest geometry of a flat uniform interface.
Using a continuum description for hydrodynamics with slip, we
derive the $(1+b/L)$ enhancement factor for that situation.
A formal generalization to other interfacially-driven phenomena is then presented.
Further, a molecular dynamics study of diffusio-osmosis in a thin slab geometry quantitatively 
conforts the picture.
We end with a brief discussion of the case of charged solutes (in particular electro-osmosis) and of  the motion of finite-size particles.\\

Consider a flat homogeneous solid surface 
$y(x,z)=0$, with an incompressible liquid of bulk viscosity $\eta_0$ in the $y>0$ space. Slip
is decsribed through the Navier 
boundary condition (BC) for the velocity field $v$,
$b\,\partial_yv_i| _{y=0}=v_i| _{y=0}$ for $i=x,z$, with $b$ the distance in the solid
at which the linearly extrapolated velocity becomes zero (see Figure 1).
In a slightly more general approach the hydrodynamic "weakness" of the interface 
shows up in a $y$-dependent viscosity $\eta(y)$, while requiring $v| _{y=0}=0$. The Navier BC is  recovered using the ansatz $\eta^{-1}(y)=\eta_0^{-1}(1+b.\delta(y))$, sketching 
slip 
in terms of a very thin vacuum layer of very low viscosity "between" the liquid and the solid.

{\em Diffusio-osmosis for a neutral solute -}
Suppose that the solution
contains a single neutral solute, at bulk concentration $c_0$,
which interacts with the wall through a short-range potential $U(y)$. 
In the dilute limit, at equilibrium the distribution of the solute is $c_{eq}(y)=c_0 \exp(-\frac{U(y)}{k_BT})$.
If a concentration gradient $dc_0/dx$ is applied along $x$ over long distances (compared to the range of the potential), equilibration of concentration and pressure is fast along $y$ 
(compared to the relaxation time of the gradient), so that 
$c \simeq c_0(x)\exp(-\frac{U(y)}{k_BT})$ and $-\partial_y p(x,y)  + c(x,y) (-\partial_y U)=0$.
This leads to the "osmotic" equilibrium $p(x,y)-k_BTc(x,y)=p_0-k_BTc_0(x)$ \cite{And91}, with $p_0$ the {constant} bulk pressure. As a consequence 
a pressure gradient along $x$ sets in (only) within the thin interfacial layer, 
$\partial_xp=k_BT\partial_x(c-c_0)$, which generates shear there through the hydrodynamic balance:
  $- \partial_x  p(x,y) + \partial_y(\eta \, \partial_y v_x)=0$.
 The fluid velocity increases accordingly through the interfacial layer to reach a finite value $v_s$,
 the "effective slip velocity" of the liquid past the surface due to the applied gradient along $x$ (Figure 1
 sketches the case of a solute attracted to the wall, $\Gamma>0$). 
 Integrating twice along $y$ and using the Navier BC:
  \begin{equation}
  v_s = - (k_BT /\eta_0) \Gamma L (1+ b/L).\frac{d c_0}{dx}
  \label{neutslip}
  \end{equation}
   where $\Gamma=\int_0^\infty \!\!dy [e^{-U(y)/k_BT} -1]$ is a length measuring the excess of solute in the vicinity of the surface ($U$ is positive and $\Gamma$ negative for depletion), and 
   $L=\Gamma^{-1}\int_0^\infty \!\!dy y [e^{-U(y)/k_BT} -1]$ measures the range of interaction of the potential. 
Equation  (\ref{neutslip}) is the classical formula of Anderson and Prieve \cite{And91}, times the amplification factor $1 + b/L$ : this quantifies how hydrodynamic slip allows 
to generate a larger "effective slip" $v_s$ away from the surface (Figure 1).
Physically $v_s$ results from the balance between viscous shear stress at the interface corrected for
slip, $\eta_0 v_s/(L+b)$, and the 
(integrated) body force within the interface layer~:
$- \frac{d}{dx}(\Gamma k_BT c_0)$.

The slip induced enhancement can actually be very large. For molecular interactions
between neutral solutes and a solid $L$ is very small, e.g. $\sim 0.3$~nm, so with
 $b\sim 20-30$~nm for water on hydrophobic substrates \cite{ref,Cot05}, the amplification factor can be up to 100 !  \\

\begin{figure}[t]
\begin{center}
\vspace{-1.5cm}
{\includegraphics[width=9cm]{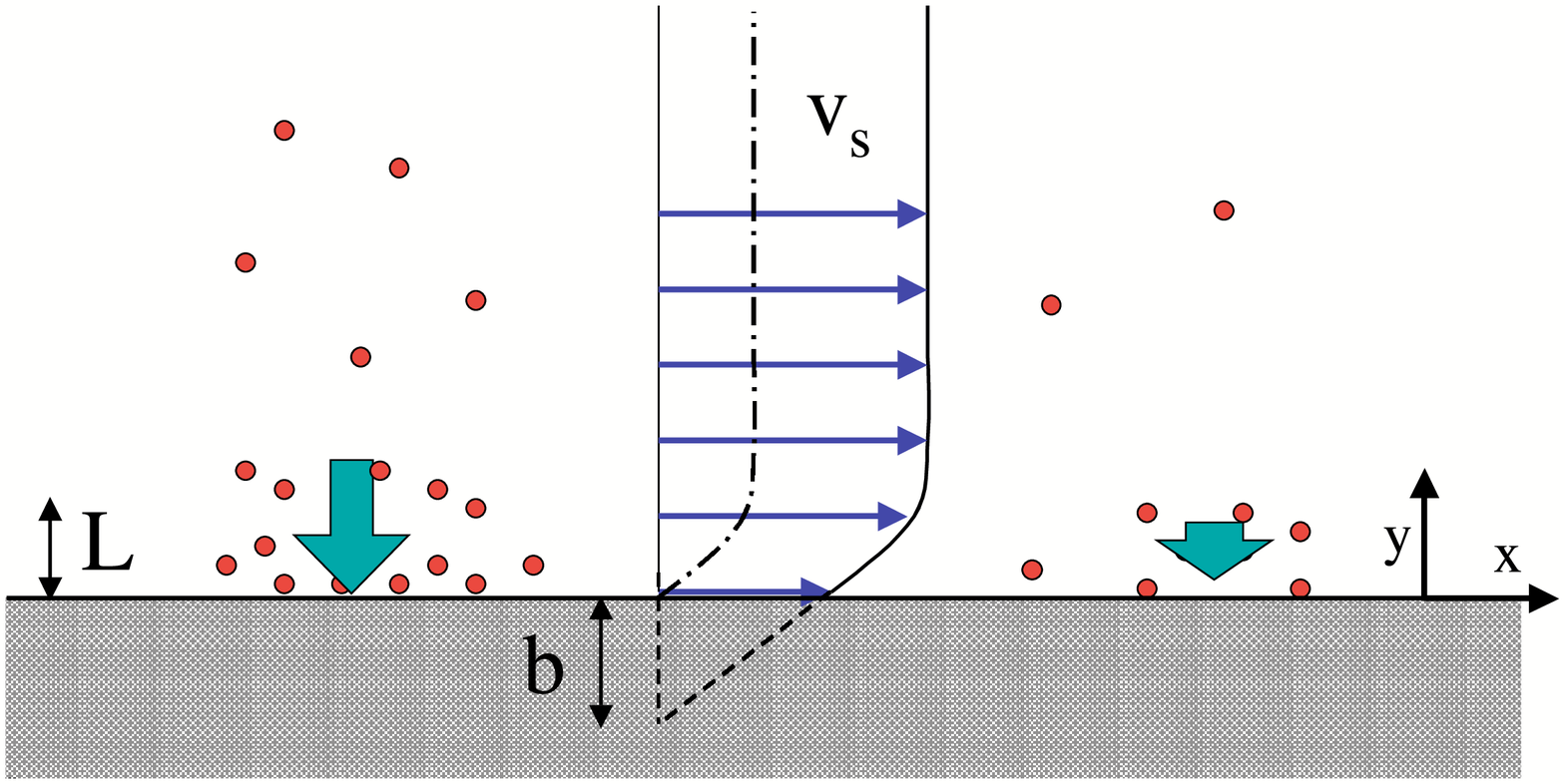}}
\end{center}
\vspace{-1.8cm}
\caption{\label{fig1} Diffusio-osmosis for a neutral solute
attracted to the solid (in grey) . The larger bulk concentration on the left results in a
higher accumulation of solute at the interface (thickness $L$) "squeezing" the fluid against the wall.
A pressure gradient results {\em in} this $L$-thick interfacial layer, which induces shear there and 
a flow opposite to the concentration gradient. In contrast with the no-slip case ($b=0$, flow depicted 
by the dot-dashed line),
hydrodynamic slip (i.e. a non-zero 
extrapolation slip length $b$) allows these stresses to generate a larger "effective" diffusiophoretic slip $v_s$
(to be distinguished from the very local slip velocity right at the wall).\\
}
\vspace{-0.7cm}
\end{figure}

{\em Formal general argument  -} 
We now generalize this result. 
For a generic interfacial structure, denote $\sigma_n$ and $\sigma_t$
the stresses normal and tangential to the interface which
develop in a thin layer close to the solid \cite{And89}. 
At equilibrium, the situation is invariant by translation along $x$,
so  $\sigma_n=\sigma_n(y)$ and $\sigma_t=\sigma_t(y)$, and 
the hydrostatic pressure $p(y)$ is determined by force balance $-\partial_y p+\partial_y \sigma_n=0$, $-\partial_x p+\partial_x \sigma_t=0$, 
yielding 
$p(y)=p_0+\sigma_n(y)$, with again $p_0$ the constant bulk pressure.

If a small far-field gradient of an observable $O$ (concentration, potential, temperature) 
is applied along $x$ then the interfacial stresses vary slowly along $x$ too. Pressure equilibration is 
fast in the $y$ direction 
and $-\partial_y p+\partial_y \sigma_n=0$ yields $p(x,y)=p_0+\sigma_n(x,y)$. 
The resulting lateral imbalance of pressure, {\it within the interfacial layer}, generates
shear along $x$ as 
described by the force balance 
$-\partial_x p+\partial_x \sigma_t+\partial_y(\eta(y)\partial_y v_x)=0$. 
Again, this generates an "effective slip" $v_s$ which reads (using $\eta(y)$ and $v(y=0)=0$):
 \begin{equation}
  v_s = - \frac{d}{dx} \left(
  \int_0^\infty \!\! \!\!dy\,\,\, \Sigma(x,y)\int_0^y \!\!\!dy' \,\eta^{-1}(y')
  \right)
  \end{equation}
with $\Sigma(x,y)=\sigma_n-\sigma_t$ the interfacial stress anisotropy. If the structure of the interface varies slowly along $x$, 
$\Sigma \simeq \Sigma_{eq}(y, O(x))$ where $O(x)$ is the "outer" value of the field $O$ outside the interfacial layer, so that the effective slip generated by $dO/dx$ is
\begin{equation}
  v_s = - {1\over \eta_0}
   \int_0^\infty \!\!\!\!\!\! dy\, \frac{\partial\Sigma_{eq}}{\partial O}(y)
    \left[
    y+ \int_0^y \!\!\!\!\!dy' 
  \,
   \frac{\eta_0-\eta(y')}{\eta(y')} 
   \right]
.
  \frac{dO}{dx} 
   \end{equation}
The integral in the bracket quantifies the specific contribution of the hydrodynamic slip.
For a slip length $b$ [using $\eta_0/\eta(y)=1+b\delta(y)$], 
we obtain our main result:
\begin{equation}
  v_s = - {1\over \eta_0}\Gamma L.[1+b/L].\frac{dO}{dx}
 \end{equation}
where $\Gamma = \int_0^\infty \!\! dy .\frac{\partial\Sigma_{eq}}{\partial O}(y)$, and $L=\Gamma^{-1}\int_0^\infty\!\! dy .y.\frac{\partial \Sigma_{eq}}{\partial O}(y)$ is a measure of the thickness of the stress-generating interface, that depends on the observable $O$ considered. 
The case of (neutral solute) diffusio-osmosis is recovered with: $\sigma_n = k_BT (c-c_0)$, $\sigma_t=0$,  $O=k_BT c_0$ and 
  $\frac{\partial\Sigma_{eq}}{\partial O}= (c_{eq}-c_0)/c_0=e^{-\frac{U(y)}{k_BT}}-1$.


 The results obtained so far rely on a continuum description of the interface hydrodynamics.
To demonstrate that the enhancement persists 
in a more realistic context, 
we turn to a slab geometry, that we will analyze using numerical simulations.\\

{\em Diffusioosmosis in a channel -}
Let us consider a channel of width $H$.
In the linear response regime
a symmetric matrix relates the fluxes (per unit length) in the $x$ direction
$(Q,J)$ to the gradients $(- \nabla \pi, -\nabla \mu)$ that generate them, with $Q$ the total flow rate, $J$ the total solute current, $\pi=p-k_BTc$ the pressure corrected for osmotic effects, and $\mu \simeq \mu_0 + k_BT \ln(c)$ the chemical potential of the solute \cite{deG,Bru04}~. 
Equivalently, diffusio-osmosis is best decribed
by the following 
matrix $M$ quantifying net transport through the channel:
\begin{equation}
\left[
\begin{array}{l}
{Q}\\ 
{ J-c_0Q} 
\end{array} 
\right]
= 
\left[
\begin{array}{lr}
{ M_{11}}&{ M_{12}}\\
{ M}_{21}&{ M_{22}}
  \end{array} 
\right] 
.
\left[ 
\begin{array}{r}
-{ \nabla}p \\
-k_BT{\nabla}c_0/c_0
\end{array} 
\right]   
\label{MatM}
\end{equation}
Onsager reciprocity relations require $M_{12}=M_{21}$ \cite{deG,Bru04}, which we explicitly
checked by solving the continuum hydrodynamic problem with a slip length $b$ on the two walls in the two situations $\nabla p \neq 0, \nabla c_0=0$
and $\nabla p = 0, \nabla c_0 \neq 0$ \cite{AB}. 
In the latter situation and 
for channels wider than the interfacial structures ($H \gg L$), we obtain as 
expected a plug-like flow driven by a concentration gradient~:
$Q = M_{12} (-k_BT\nabla c_0/c_0)= H. v_s$ 
with $v_s$ the slip velocity given in Eq. (\ref{neutslip}), and $M_{12}= {H c_0\over \eta_0}\Gamma (L + b)$.
For thinner (nano)
channels, the overlap between the interfacial layers must be taken into account \cite{AB}.\\

 {\em Numerical Simulations  - }
 We then conduct Molecular Dynamics simulations of a 
fluid system  composed of solvent+solute particles, confined between two parallel solid walls composed of individual "solid" particles fixed on a fcc lattice \cite{Plimpton}. Interactions between the three types of particles are
of Lennard-Jones type, 
$U_{\alpha\beta}(r)=4 \epsilon \left[ \left({\sigma \over r}\right)^{12} - u_{\alpha\beta}  \left({\sigma \over r}\right)^{6} \right]$
with identical interaction energy $\epsilon$ and molecular diameters $\sigma$ ($\alpha,\beta\in$ \{solute,solvent,walls\}). Tuning the 
parameters $u_{\alpha,\beta}$ we can vary (i)  
the wettability of the solvent on the wall by tuning $u_{\rm solvent,wall}$,
 and (ii) the
relative attraction or depletion of the solute to that wall
(by tuning  $u_{\rm wall,solute}$ for a fixed $u_{\rm wall,solvent}$).
Periodic boundary conditions are used along $x$ and $z$
(box size $l_x=l_z=16 \sigma$), and the inter-wall distance is $l_y= H=20.8 \sigma$.
Temperature is kept
constant by applying a Hoover thermostat 
to the $z$ degrees of freedom
({\it i.e.} 
perpendicular to flow and confinement).
Solvent density is $\rho_f \sigma^3\sim 0.9$, and bulk solute concentration 
$c_0 \sigma^3\sim 0.02$.
Rather long runs ($\sim 5.10^6$ timesteps) are performed to obtain good statistics.

To determine the cross coefficient $M_{12}=M_{21}$,
the most efficient route is to apply an 
external volume force, $f_0=-\nabla p$, to the fluid in the $x$ direction, so as to generate a pressure-driven flow.
We then measure the solute {\it excess current}, $J-c_0 Q$, associated with the convective motion of the solute \cite{note}, and obtain $M_{21}=(J-c_0 Q)/(-\nabla p)$, according to eq. (\ref{MatM})
(we check linearity of the reponse to the external force). Eventually we extract the adsorption length $\Gamma$ from the equilibrium solute density profile $c(y)$ as 
$\Gamma={1\over 2}\int_{slab} dy [c(y)/c_0 -1]$.
To narrow our exploration, we focus on the ideal solution of solvent and solute molecules identical but for their interactions with the walls. We take $u_{\rm solvent,solvent}=u_{\rm solute,solute}=u_{\rm solvent,solute}=1.2$, and consider three solvent-wall situations
$u_{\rm wall,solvent}=0.3, 0.5, 1.0$ (going from non-wetting to wetting) and
solute-wall interactions $u_{\rm wall,solute}$ in the range $[0.1,1.1]$. 
In all cases, the hydrodynamic velocity profiles are parabolic, which allows us to
extract the viscosity $\eta$ and
the slip length $b$ \cite{BB}.
In agreement with previous work \cite{BB} and experimental results, 
slip is significant for a non-wetting solvent
($b\sim 20-40 \sigma$ for $u_{\rm wall,solvent} \sim 0.3-0.5$), and negligible
for a wetting solvent ($b\leq \sigma$ for $u_{\rm wall,solvent} = 1$).

\begin{figure}[t]
\begin{center}
\vspace{-0.8cm}
\includegraphics[width=8cm]{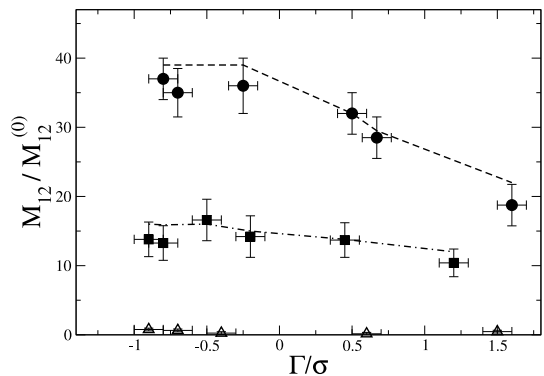}
\vspace{-0.5cm}
\caption{Cross coefficient $M_{12}$ for a slab geometry (normalized by a no-slip reference $M_{12}^{(0)}={c_0\over \eta} \sigma H\Gamma $),
against normalized interfacial enrichment in solute $\Gamma/\sigma$, for wetting to non-wetting solvents $u_{\rm wall,solvent}=1.0$ ($\scriptscriptstyle\triangle$), $0.5$ ($\scriptscriptstyle\blacksquare$),$0.3$ 
($\bullet$), and $u_{\rm wall,solute}$ in the range $[0.1-1.1]$. The dashed lines
are the theoretical prediction $M_{12}/M_{12}^{(0)}=(L+b)/\sigma \simeq b/\sigma$
using the measured slip length $b$ which is significant for the non-wetting cases: 
$b\sim  12-16\sigma$ ($\scriptscriptstyle\blacksquare$)
and $b\sim 20-40\sigma$ ($\bullet$). 
For large
positive adsorption, the enhancement decreases as does the slip length, because the adsorbed solute
increases the fluid/solid wetting. 
}
\label{fig2}
\end{center} 
\vspace{-.8cm}
\end{figure}

Fig. \ref{fig2} displays the outcome of our simulations for the cross coefficient $M_{21}=M_{12}$, normalized by a value $M_{12}^{(0)}={c_0\over \eta} \sigma H\Gamma $, which corresponds to a
reference situation with a no-slip BC and fixed $L=\sigma$.  
In line with our theoretical arguments, $M_{12}$ is strongly enhanced  -- $M_{12}/M_{12}^{(0)}$
up to $40$ here--
for non-wetting solvents  ($u_{\rm wall,solvent}=0.3, 0.5$, top data), i.e. 
systems with slip lengths of a few tens 
of molecular diameters ({\it i.e.} roughly $\sim 5-10 nm$).
On the other hand, $M_{12}$ is much smaller  ($M_{12}Ê\sim M_{12}^{(0)}$) for a wetting solvent ($u_{\rm wall,solvent}=1$, bottom data), associated with a no-slip
BC. 

More quantitatively our MD results compare succesfully to the
theoretical prediction rewritten as $M_{12}Ê/M_{12}^{(0)}= (L+b)/\sigma\simeq b/\sigma$ (since $L\lesssim\sigma \ll b$), see Figure \ref{fig2}, provided we use the slip length $b$ extracted from the simulation for each case. In particular,
the amplification decreases for large positive adsorptions $\Gamma>0$, due to the decrease of the slip length~: accumulation of "wetting" solute 
 at the solid-liquid interface reduces the effective "solvophobicity" \cite{note2}. 
 For a depleted solute, ($\Gamma<0$), this "saturation" effect is essentially absent 
($b$ is nearly constant) allowing for large enhancements.\\




{\em Electrolyte solutions: electro- and diffusio-osmosis -}
For charged solutes, the interfacial structure
is the electrical double layer, of typical thickness the Debye length $\kappa^{-1}$ \cite{Hun},
usually in the nanometer range ($1-30$~nm), and we anticipate $L\sim \kappa^{-1}$.
The enhancement of interfacially driven phenomena (electro-osmosis, diffusio-osmosis, thermophoresis) over solvophobic surfaces ($b$ in the $20-30$~nm range) should thus be somewhat smaller than for the neutral solute case, but still significant.

As a check of $L\sim \kappa^{-1}$, we incorporate a finite slip length $b$ in the usual description of these phenomena \cite{Hun}, and compute the enhancement factor $(1+b/L)$ in equation (4) \cite{AB}.
For electro-osmosis 
$L=\frac   {-\phi_{eq}} {d\phi_{eq}/dy}|_{y=0}$, with $\phi_{eq}(y)$ the equilibrium electrostatic potential in the double layer, so
for weakly charged surfaces $L\simeq \kappa ^{-1}$ in agreement with \cite{Chu,Sto04,Jol04}. 
For diffusio-osmosis and a 1:1 electrolyte, we obtain a more complex formula, with $L\simeq \kappa^{-1}/2$ for  weakly charged surfaces.\\


{\em Transport of particles -}
All the above applies to the reciprocal motion
of particles in concentration or potential fields, 
in a way that can be quite directly quantified provided the surface is locally
flat and homogeneous at the $L$ scale following \cite{Hun,And89,And91}.
Classically for interfacial driven effects, considering finite-size
objects such as a spherical particle of radius $a \gg L$, allows one to discuss
the possible feed-back of the generated flow on the interfacial structure 
where it originates \cite{And91}.
We compute here the diffusiophoresis of such a sphere 
 generated by a steady background gradient of neutral solute,
adapting the classical no-slip analysis of \cite{And91}. Including hydrodynamic slip (non-zero $b$) enhances the surface/liquid effective slip as described by (1), but also the convection of solute in the interfacial region, which affects the steady-state concentration field of the solute (diffusion coefficient $D$) around the particle. We find the velocity $U$ of the particle in a solute gradient $\nabla c_0$ to be:
\begin{equation}
{\bf U}= \frac{k_BT}{\eta_0}.L\Gamma.\frac{1+b/L}{1+ (1+(\nu +b/L)Pe)(\Gamma/a)}.{\mathbf \nabla}c_0
\end{equation}
with $Pe=(k_BT/D\eta_0)L\Gamma c_0$ and $\nu$ a dimensionless quantity of order $1$ defined in \cite{And91} that depends on the exact shape of the potential. 
For moderate values of  $\Gamma\ll a$, the usual slip enhancement factor $(1+b/L)$ prevails,
and for $b\gg L$ the formula reads $U \simeq (k_BT/\eta_0).b\Gamma.{ \nabla}c_0$.
For ${\nabla}c_0\sim 10^{-3} {\rm mol/cm^4}$, $\Gamma\sim L\sim  \AA$,
and $b\sim 30 {\rm nm}$, this leads to 
$ U \gtrsim \mu {\rm m/s}$ (in contrast to $\sim 5 {\rm nm/s}$ for the no slip case !), comparable to experimental observations
of chemical self-propulsion \cite{Pax}.
For smaller particles or stronger solute adsorption ($\Gamma/a\gg 1$), the effect of slip saturates for large $b/L$
(the large generated flow "erases" partly the original interfacial gradients),
with a maximal velocity  $U_{max}= (k_BT/\eta_0).\frac{La}{Pe}.{ \nabla}c_0$ independent of $b$. \\


{\em Conclusion -}
Hydrodynamic slip can very significantly
enhance many interfacially-driven
phenomena on smooth "solvophobic" surfaces. 
This is of relevance for the transport of fluids in small channels,
and of particles in solutions. A related target is the modelling and engineering 
of the self-transport of chemically-driven swimmers, 
for which the hydrophobicity of the surface 
is thought to play an important role \cite{Pax}. 
Further study is necessary to go beyond the model smooth surfaces considered here, so as to 
assess the effect of topographic or chemical heterogeneities at various scales (e.g. roughness can potentially either increase or decrease slip effects in channels depending on whether or not it leads to gas entrapment).



\begin{thebibliography}{13}
\bibitem{Squ05} T.M. Squires, S.R. Quake, Rev. Mod. Phys. {\bf 77}, 977-1026 (2005).
\bibitem{Sto04} H. Stone, A. Stroock, A. Ajdari, Ann. Rev. Fluid. Mech. {\bf 36}, 381 (2004).
\bibitem{Hun} R.J. Hunter, {\em Foundations of Colloid Science}, 
Oxford University Press, New York 1991.
\bibitem{And89} J.L. Anderson, Ann. Rev. Fluid. Mech. {\bf 21}, 61-99 (1989).
\bibitem{ref} E. Lauga, M. Brenner, H. Stone, Handbook of Experimental Fluid Dynamics (Springer, 2006).
\bibitem{Cot05} C. Cottin-Bizonne, B. Cross, A. Steinberger, E. Charlaix
Phys. Rev. Lett.,{\bf 94}, 056102 (2005).
\bibitem{Roth} J. Ou, B. Perot, J.P. Rothstein, Phys. Fluids {\bf 17} 4735 (2004).
\bibitem{Jol04} L. Joly, C. Ybert, E. Trizac, L. Bocquet, Phys. Rev. Lett., {\bf 93}, 257805 (2004).
\bibitem{Chu} N.V. Churaev, J. Ralston, I.P. Sergeeva, V.D. Sobolev, Adv. Coll. Int. Sci. {\bf 96}Ê265 (2002).
\bibitem{Lam} P.E. Lammert, R. Bruinsma, J. Prost, {\em J. Theor. Biol.}, {\bf 178}, 387-391 (1996).
\bibitem{Gol05} R. Golestanian, T. Liverpool, A. Ajdari, Phys.Rev.Lett., {\bf 94}, 220801 (2005) . 
\bibitem{Pax}  W.F. Paxton, A. Sen, T. E. Mallouk, {\em Chem. Eur. J.}, {\bf 11}, 6462-6470 (2005).
\bibitem{And91}  J.L. Anderson, D. Prieve, Langmuir, {\bf 7}, 403-406 (1991).
\bibitem{deG} 
De Groot S.R., Mazur P., {\em Non-equilibrium Thermodynamics},
North Holland, Amsterdam (1969).
\bibitem{Bru04} E. Brunet, A. Ajdari, Phys.Rev.E, {\bf 69}, 016306 (2004).
\bibitem{AB} A. Ajdari, L. Bocquet, in preparation.
\bibitem{Plimpton} We have used the MD code LAMMPS 2001, written by Steve Plimpton (Sandia Labs.)
\bibitem{note} $c_0$ is measured in the middle of the cell,
where the solute density profile is flat; $J$ and $Q$ are measured resp. from
the averaged solute current and solvent velocity profile.
\bibitem{BB} J.-L. Barrat, L. Bocquet, {\it Phys. Rev. Lett.} {\bf 82} 4671 (1999).

\bibitem{note2} A simple empirical law quantitatively fits the numerical results for $b$: 
$1/b=(1-x_0)/b_0+x_0/b_1$, with $x_0=c_0/\rho_f$ the solute volume fraction 
 and $b_0$ (resp. $b_1$)  the slip length for a pure solvent (resp. solute) system. 
\end{thebibliography}
\end{document}